\documentclass[twocolumn,showpacs,preprintnumbers]{revtex4}%
\usepackage{amssymb}
\usepackage{amsmath}
\usepackage{graphicx}
\usepackage{dcolumn}
\usepackage{bm}
\usepackage{amsfonts}%
\begin{document}
\title{Multiphoton resonances in nitrogen-vacancy defects in diamond}
\author{Sergei Masis}
\author{Nir Alfasi}
\author{Roei Levi}
\author{Oleg Shtempluck}
\author{Eyal Buks}
\affiliation{Andrew and Erna Viterbi Department of Electrical Engineering, Technion - Haifa 32000, Israel}
\date{\today }

\begin{abstract}
Dense ensembles of nitrogen vacancy (NV) centers in diamond are of interest  for various applications including magnetometry, masers, hyperpolarization and quantum memory. All of the applications above may benefit from a non-linear response of the ensemble, and hence multiphoton processes are of importance. We study an enhancement of the NV ensemble multiphoton response due to coupling to a superconducting cavity or to an ensemble of Nitrogen 14 substitutional defects (P1). In the latter case, the increased NV sensitivity allowed us  to probe the P1 hyperfine splitting. As an example of an application, an increased responsivity to magnetic field is demonstrated.
\end{abstract}
\maketitle



%It is always \today, today,
%but any date may be explicitly specified

%PACS, the Physics and Astronomy
%Classification Scheme.

\section{Introduction}

A two level system (TLS) is perhaps the most extreme manifestation of
nonlinear response. Systems composed of TLSs and other elements exhibit a
variety of nonlinear dynamical effects including multi-photon resonances (MPR)
\cite{Shirley_B979,Berns_150502,Tycko_1761,Faisal_theory}, frequency mixing
\cite{Childress_033839,Chen_167401,Mamin_030803}, fluorescence
\cite{Mollow_525,Freedhoff_474}, dynamical instabilities
\cite{Weber_681,Armen_063801}, suppression of tunneling
\cite{Grossmann_516,Lignier_220403} and breakdown of the rotating wave
approximation \cite{Fuchs_1181193}.

Here we study nonlinear response of an ensemble of nitrogen-vacancy (NV)
defects in diamond \cite{Doherty_1}. Two mechanisms that allow the enhancement
of MPR are explored. The first one is based on an
electromagnetic cavity mode that is coupled to the spin ensemble
\cite{Zhu2011,Kubo_140502,Kubo_220501,Amsuss_060502,Schuster_140501,Sandner_053806,Grezes_021049,Alfasi_063808}%
. The second one is attributed to hyperfine splitting \cite{Alvarez_8456} of P1 defects 
\cite{Kamp_045204,Sushkov_197601,Belthangady_157601} and their dipolar coupling to the negatively charged NV defects ($\text{NV}^{-}$).

The $\text{NV}^{-}$ defect has a spin triplet
ground state \cite{Doherty_205203} having relatively long coherence time
\cite{Balasubramanian_383}. The NV$^{-}$ spin state can be initiated via the
process of optically-induced spin polarization (OISP)
\cite{Robledo_025013,Redman_3420} and can be measured using the technique of
optical detection of magnetic resonance (ODMR)
\cite{Shin_124519,Chapman_190,Gruber_2012}. These properties facilitate a
variety of applications including magnetometry
\cite{Maze_644,Acosta_070801,Balasubramanian_648,Wolf_041001,Mamin_557,Pelliccione_700,Rondin_2279,Sushkov_197601}%
, sensing \cite{Acosta_070801,Dolde_459,Balasubramanian_383,Jelezko_076401}
and quantum information processing \cite{Maurer_1283,Cai_093030}.

Dipolar coupling between NV$^{-}$ and other spin species in diamond gives rise
to intriguing effects including hyperpolarization
\cite{Takahashi_047601,Fischer_057601,Fischer_125207,Wang_1940} and
cross-relaxation \cite{Solomon_559,Belthangady_157601,Loretz_064413}, and can
be exploited for optical detection of spin defects in diamond other than NV$^{-}$
\cite{Simanovskaia_224106,Kamp_045204,Clevenson_021401,Wang_4135,Hall_10211,Purser_1802_09635,Alfasi_2018}.

The process of cross-polarization between NV$^-$ and P1 defects plays an important role in the MPR mechanism. In general, the efficiency of cross polarization depends on the rate of a competing effect of thermal polarization, which is characterized by the longitudinal spin relaxation rate. At cryogenic temperatures the thermal polarization rate can be significantly reduced, and consequently the efficiently of cross-polarization is enhanced.
\section{Low Magnetic Field ODMR}

\begin{figure}
[ptb]
\begin{center}
\includegraphics[
width=\columnwidth
]%
{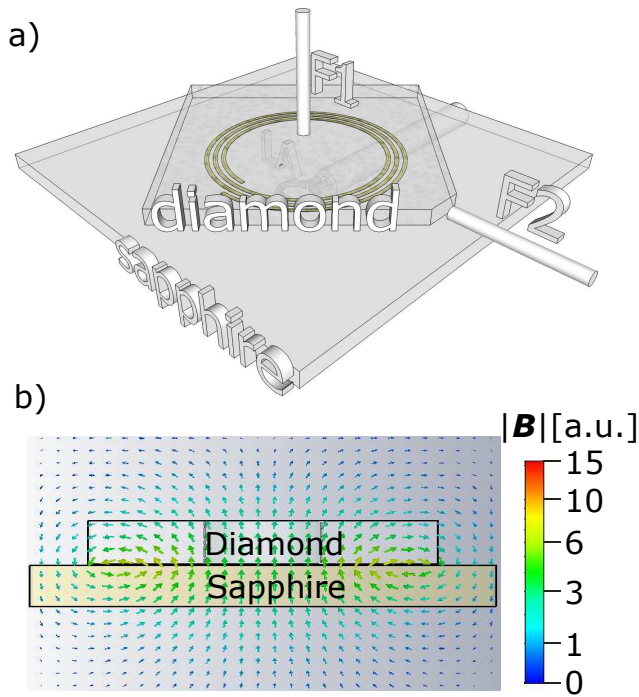}%
\caption{(a) Experimental setup. The diamond is glued on the top of the spiral resonator, and two multimode optical fibers F1 and F2 are attached to the diamond top and side faces correspondingly. A $532\operatorname{ nm}$ wavelength laser is introduced from one of the  fibers, and the photoluminescence is collected from the other, providing a geometrical filtering of the laser light. A microwave loop antenna is placed below the Sapphire at a location optimizing the resonator coupling. (b) Numerical simulation \cite{CST} of the spiral fundamental mode magnetic field distribution.}%
\label{Fig_Setup}%
\end{center}
\end{figure}

A spiral resonator \cite{Kurter_709} made of $500\operatorname{nm}/10\operatorname{nm}$ thick Niobium/Aluminum with the inner radius of $0.7\operatorname{mm}$ and line width and spacing of $20\operatorname{\mu m}$ is fabricated on a Sapphire substrate. Type Ib [110] diamond is irradiated with $2.8\operatorname{MeV}$ electrons at a doze of $8\times10^{18}\operatorname{e/cm}^2$, annealed for $2$ hours at $900\operatorname{C}^\circ$ and acid cleaned.
The samples assembly (see Fig. \ref{Fig_Setup}) is placed at a cryostat with base temperature of $3.6\operatorname{K}$ and mechanically aligned along the magnetic field of an external superconducting solenoid.
The photoluminescence light passes through an array of filters and is collected by a photodiode.  
A microwave synthesizer is connected directly to a loop antenna (shortened end of a coaxial cable) mounted below the sapphire substrate, and the signal amplitude is $100\%$ modulated with a low frequency sine wave. The same wave is used for the photodiode signal demodulation by a lock-in amplifier. Microwave reflection measurements of the resonator yield resonance frequency $\omega_\mathrm{c}=2\pi\times276\operatorname{ MHz}$, unloaded %190325
quality factor $Q=96$ and critical temperature $T_\text{c}=7\operatorname{K}$. The rather low $Q$ might be explained by the proximity to irradiated diamond. The  coupling coefficient $g$ between the resonator and the $\text{NV}^{-}$ ensemble is given by \cite{Alfasi_063808}
\begin{equation}
g^{2}=\frac
	{\gamma_{\mathrm{e}}^{2}\mu_{0}\hbar\omega_{\mathrm{c}}
		\int d\mathbf{r} n_{\mathrm{S}} P_{z}
		\left\vert \mathbf{B}_{\mathrm{c}}\right\vert^{2}
		\sin^{2}\varphi }
	{\int d\mathbf{r}{}
		\left\vert\mathbf{B}_{\mathrm{c}}\right\vert^{2}
	}
	,
	\label{g_s^2}%
\end{equation}
where $P_{z}\approx0.15$ \cite{Alfasi_2018} is the spin polarization, $n_{\mathrm{S}}$ is the NV$^-$ ensemble number density, $\varphi$ is the angle between the NV$^-$ axis and the cavity magnetic field $\mathbf{B}_{\mathrm{c}}$, $\mu _{0}$ is the free space permeability and $\gamma _{\mathrm{e}}=2\pi \times 28.03\operatorname{GHz}\operatorname{T}^{-1}$ is the electron spin gyromagnetic ratio. Assuming constant $P_z$ throughout the diamond, $g$ is readily calculated by means of numerical simulation [see Fig.\ref{Fig_Setup}(b)] to be $g=8\operatorname{MHz}$.
%sqrt((2*pi*28.03e9)^2*1.257e-6*1.0545e-34*276e6*2*pi*0.15*0.2/(3e-3*3e-3*0.5e-3)*(3e17*0.3*0.3*0.05))=8010500.0401
%

%TCIMACRO{\FRAME{ftbpFU}{3.4546in}{1.8468in}{0pt}{\Qcb{Experimental setup.}%
%}{\Qlb{Fig_Setup}}{fig_setup.eps}{\special{ language "Scientific Word";
%type "GRAPHIC";  maintain-aspect-ratio TRUE;  display "ICON";
%valid_file "F";  width 3.4546in;  height 1.8468in;  depth 0pt;
%original-width 15.1709in;  original-height 8.055in;  cropleft "0";
%filename 'Fig_Setup.eps';file-properties "XNPEU";}} }%
%BeginExpansion
%EndExpansion

\begin{figure}
[ptb]
\begin{center}
\includegraphics
%[
%height=2.5949in,
%width=3.4537in
%]%
[width=\columnwidth]
{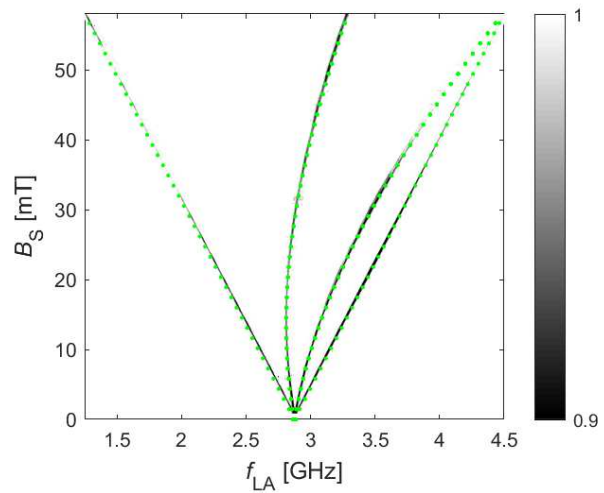}%
%ParamSet 100, data 20180202-All_Data_Combined 
\caption{Low magnetic field ODMR. The overlaid green dotted lines are calculated by
diagonalizing the NV$^{-}$ spin Hamiltonian (\ref{H NV-}). The two nearly straight diagonal curves (leftmost and rightmost) correspond to the NV axis vector nearly parallel to the magnetic field, while the two remaining curves in the middle - to the nearly degenerate three other possible orientations of the NV axis vectors. }%
\label{Fig_ODMR}%
\end{center}
\end{figure}

ODMR as a function of magnetic field and frequency is shown in Fig.
\ref{Fig_ODMR}. The dotted lines in Fig. \ref{Fig_ODMR} are
calculated by numerically diagonalizing the NV$^{-}$ ground state spin triplet
Hamiltonian, which is given by
\cite{Ovartchaiyapong_1403_4173,MacQuarrie_227602}
\begin{equation}
\frac{\mathcal{H}_{\mathrm{NV}}}{\hbar}=\frac{DS_{z}^{2}}{\hbar^{2}}%
+\frac{E\left(  S_{+}^{2}+S_{-}^{2}\right)  }{2\hbar^{2}}-\frac{\gamma
_{\mathrm{e}}\mathbf{B}\cdot\mathbf{S}}{\hbar}\;,\label{H NV-}%
\end{equation}
where $\mathbf{S}=\left(  S_{x},S_{y},S_{z}\right)  $ is a vector spin $S=1$
operator, the raising $S_{+}$ and lowering $S_{-}$ operators are defined by
$S_{\pm}=S_{x}\pm iS_{y}$, the zero field splitting induced by spin-spin
interaction $D$ is given by $D=2\pi\times2.87%
%TCIMACRO{\unit{GHz}}%
%BeginExpansion
\operatorname{GHz}%
%EndExpansion
$, the strain-induced splitting $E$ is about $2\pi\times10%
%TCIMACRO{\unit{MHz}}%
%BeginExpansion
\operatorname{MHz}%
%EndExpansion
$ for our sample and $\mathbf{B}$\ is the externally applied magnetic field. The field $\mathbf{B}$ has two contributions $\mathbf{B}=\mathbf{B}%
_{\mathrm{S}}+\mathbf{B}_{\mathrm{L}}$, where $\mathbf{B}_{\mathrm{S}}$ ($\mathbf{B}_{\mathrm{L}}$) is the stationary (alternating) field generated by the solenoid (the loop antenna) and is nearly parallel to the lattice direction $\left[  111\right]  $ ($\left[  1\bar{1}0\right]  $).

In a single crystal diamond the NV centers have four different possible
orientations. When hyperfine interaction is disregarded each orientation gives
rise to a pair of angular resonance frequencies $\omega_{\pm}$, corresponding
to the transitions between the spin state with magnetic quantum number $0$ and
the spin state with magnetic quantum number $\pm1$. The dotted line in Fig.
\ref{Fig_ODMR} having the smallest (largest) frequency for any given magnetic
field corresponds to the angular frequency $\omega_{-}$ ($\omega_{+}$) of the
NV$^{-}$ defects having axis in the $\left[  111\right]  $ lattice direction. The other two dotted
lines represent the resonances due to the unparallel NV$^{-}$ defects having
axis in the lattice directions $\left[  \bar{1}11\right]  $, $\left[  1\bar
{1}1\right]  $ or $\left[  11\bar{1}\right]  $. 

%TCIMACRO{\FRAME{ftbpFU}{3.4537in}{2.5949in}{0pt}{\Qcb{Low magnetic field ODMR.
%The overlaid dotted lines are calculated by diagonalizing the NV$^{-}$ spin
%Hamiltonian (\ref{H NV-}).}}{\Qlb{Fig_ODMR}}{fig_odmr.eps}%
%maintain-aspect-ratio TRUE;  display "ICON";  valid_file "F";
%width 3.4537in;  height 2.5949in;  depth 0pt;  original-width 8.5365in;
%original-height 6.3945in;  cropleft "0";  croptop "1";  cropright "1";
%cropbottom "0";  filename 'Fig_ODMR.eps';file-properties "XNPEU";}} }%
%BeginExpansion
%EndExpansion

\section{Near the Level Anti Crossing}

\begin{figure}
[ptb]
\begin{center}
\includegraphics[
height=2.5949in,
width=3.4537in
]%
{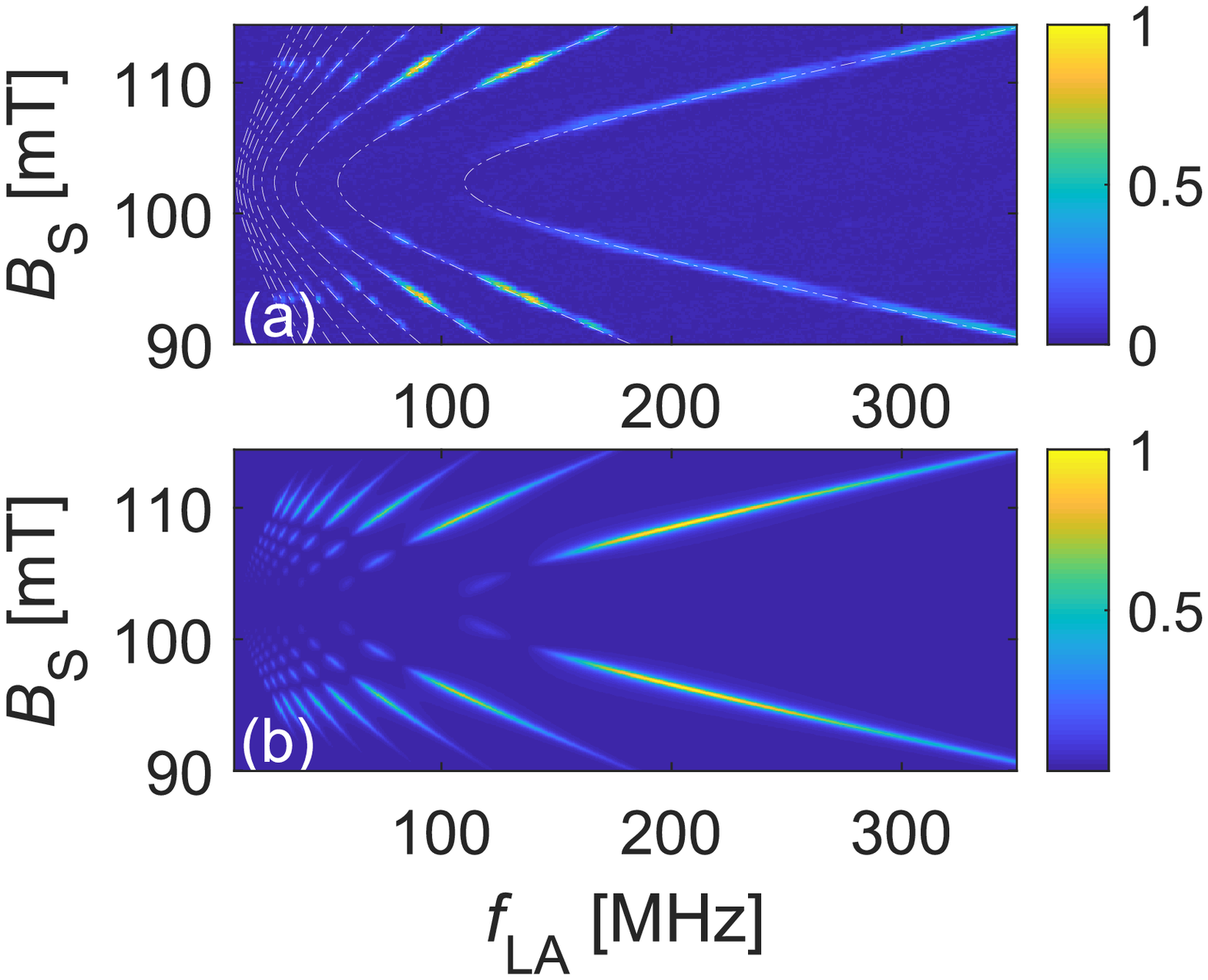}%
\caption{ODMR with low laser power. (a) The normalized ODMR signal as a
function of driving frequency $f_{\mathrm{LA}}$ and magnetic field
$B_{\mathrm{S}}$. The overlaid hyperbolas are calculated according to Eq.
(\ref{hyperbolas}). (b) The normalized steady state polarization
$P_{z}/P_{z,\mathrm{s}}$ is calculated according the Eq. (\ref{P_z SS=}) with
the following parameters $\omega_{\mathrm{c}}=2\pi\times276\operatorname{MHz}
$, $\gamma_{\mathrm{c}}=2.87\operatorname{MHz}$, $\gamma_{1}=20\operatorname{Hz}$ and $\gamma_{2}=30\operatorname{MHz}$. In terms of the
parameter $\eta$ the coupling coefficient $\beta_{\Delta}$\ is expressed as
$\beta_{\Delta}=\beta_{\Delta0}\eta/\sqrt{1+\eta^{2}}$, where $\beta_{\Delta
0}=10$.}%
\label{Fig_LowL}%
\end{center}
\end{figure}

Let $\omega_{\mathrm{a}}$ denote the angular frequency $\omega_{-}$
corresponding to the NV$^{-}$ defects having axis in the $\left[  111\right]
$ lattice direction. Consider the case where the magnitude $B_{\mathrm{S}}$ of
the solenoid field $\mathbf{B}_{\mathrm{S}}$ is tuned close to the value
$D/\gamma_{\mathrm{e}}=102\operatorname{mT}$. In the vicinity of this level anti-crossing point (LAC) the
angular frequency $\omega_{\mathrm{a}}$ is approximately given by
$\omega_{\mathrm{a}}=\omega_{\mathrm{a}0}\sqrt{1+\eta^{2}}$, where
$\omega_{\mathrm{a}0}=\sqrt{2}D\theta_{\mathrm{S}}$ is the lowest value of the
angular frequency $\omega_{\mathrm{a}}$, $\theta_{\mathrm{S}}\ll1$ is the
angle between $\mathbf{B}_{\mathrm{S}}$ and the lattice direction $\left[
111\right]  $ ($\theta_{\mathrm{S}}=1.5^{\circ}$ and $\omega_{\mathrm{a}%
0}/2\pi=110%
%TCIMACRO{\unit{MHz}}%
%BeginExpansion
\operatorname{MHz}%
%EndExpansion
$ for the data shown in Figs. \ref{Fig_ODMR}, \ref{Fig_LowL} and
\ref{Fig_HighL}), and the dimensionless detuning $\eta$ is given by
$\eta=\gamma_{\mathrm{e}}\delta B_{\mathrm{S}}/\omega_{\mathrm{a}0}$, where
$\delta B_{\mathrm{S}}=B_{\mathrm{S}}-D/\gamma_{\mathrm{e}}$.

Measured ODMR near the LAC vs.
magnetic field $B_{\mathrm{S}}$ and driving frequency $f_{\mathrm{LA}}%
=\omega_{\mathrm{LA}}/2\pi$ of the signal injected into the loop antenna is
seen in Fig. \ref{Fig_LowL}(a). 
%(data 20180104-2039)
The overlaid white dotted lines
are hyperbolas calculated according to $f_{\mathrm{LA}}=f_{l}$, where the
frequency of $l$'th hyperbola $f_{l}$ is given by%
\begin{equation}
f_{l}=\omega_{\mathrm{a}}/2\pi l=\omega_{\mathrm{a}0}%
\sqrt{1+\eta^{2}}/2\pi l\;, \label{hyperbolas}%
\end{equation}
where $l$ is an integer from $1$ to $10$.
As can be seen from Fig. \ref{Fig_LowL}, along the $l$'th hyperbola the largest signal is obtained when the driving frequency is tuned close to $\omega_\text{c}/2\pi l$,  where $\omega_\text{c}/2\pi=276\operatorname{MHz}$ is the cavity resonance frequency. This suggests that the spin MPR are enhanced due to the interaction with the cavity mode.
%

%TCIMACRO{\FRAME{ftbpFU}{3.4537in}{2.5949in}{0pt}{\Qcb{ODMR with low laser
%power. (a) The normalized ODMR signal as a function of driving frequency
%$f_{\mathrm{LA}}$ and magnetic field $B_{\mathrm{S}}$. The overlaid hyperbolas
%are calculated according to Eq. (\ref{hyperbolas}). (b) The normalized steady
%state polarization $P_{z}/P_{z,\mathrm{s}}$ is calculated according the Eq.
%(\ref{P_z SS=}) with the following parameters $\omega_{\mathrm{c}}=2\pi
%\times276\unit{MHz}$, $\gamma_{\mathrm{c}}=10^{4}\unit{Hz}$, $\gamma
%_{1}=10^{4}\unit{Hz}$ and $\gamma_{2}=10^{7}\unit{Hz}$. In terms of the
%parameter $\eta$ the coupling coefficient $\beta_{\Delta}$\ is expressed as
%$\beta_{\Delta}=\beta_{\Delta0}\eta/\sqrt{1+\eta^{2}}$, where $\beta_{\Delta
%0}=10$.}}{\Qlb{Fig_LowL}}{fig_lowl.eps}{\special{ language "Scientific Word";
%type "GRAPHIC";  maintain-aspect-ratio TRUE;  display "ICON";
%valid_file "F";  width 3.4537in;  height 2.5949in;  depth 0pt;
%original-width 8.5365in;  original-height 6.3945in;  cropleft "0";
%filename 'Fig_LowL.eps';file-properties "XNPEU";}} }%
%BeginExpansion
%EndExpansion

\section{Cavity Superharmonic Resonances}

The effect of the coupled cavity mode on the spin MPR is discussed in Appendix \ref{appA}.  
The theoretical model presented in the appendix describes the interplay
between two mechanisms. The first one is frequency mixing between transverse
and longitudinal spin driving. Near the avoided crossing point the NV$^{-}$
spin states with magnetic quantum numbers $-1$ and $0$ are mixed, and
consequently the amplitudes of transverse and longitudinal driving become
strongly dependent on detuning from the avoided crossing point (even when
the external driving is kept unchanged). The highly nonlinear nature of the
first mechanism results in the generation of harmonics of the externally
applied driving frequency. The second mechanism is cavity resonance
enhancement, which becomes efficient when one of the generated harmonics
coincides with the cavity resonance band. Under appropriate conditions this may
give rise to a pronounce cavity-assisted multi-photon resonance.

Consider the case where the frequency of excitation injected into the loop
antenna is tuned close to the $l$'th superharmonic resonance, i.e.
$\omega_{\mathrm{a}}\simeq l\omega_{\mathrm{LA}}$, where $l$ is an integer. 
%Intuitively, the NV spins ensemble rotates in all the harmonics of $\omega_{\mathrm{LA}}$, exciting the spiral resonator, that in turn reduces the polarization of the resonant NV, causing an increase in the absolute ODMR signal. More formally, 
In that region, the relative change $P_{z}/P_{z,\mathrm{s}}$ in spin polarization
in the NV$^{-}$ triplet ground state is found to be given by [see Eq.
(\ref{P_z SS}) in appendix \ref{appA}]%
\begin{equation}
\frac{P_{z}}{P_{z,\mathrm{s}}}=1-\frac{\beta_{\Delta}^{2}\left\vert
\zeta\right\vert ^{2}}{1+\beta_{\Delta}^{2}\left\vert \zeta\right\vert
^{2}+\beta_{\mathrm{a}l}^{2}}\;,\label{P_z SS=}%
\end{equation}
where [see Eq. (\ref{zeta_a})]%
\begin{equation}
\zeta=J_{l}\left(  \frac{\omega_{\mathrm{b}}}{\omega_{\mathrm{L}}}\right)
\left(  1+\frac{\kappa J_{0}^{2}\left(  \frac{\omega_{\mathrm{b}}}%
{\omega_{\mathrm{L}}}\right)  P_{z,\mathrm{s}}}{\left(  1+i\beta_{\mathrm{c}%
l}\right)  \left(  1+i\beta_{\mathrm{a}l}\right)  }\right)  \;,
\end{equation}
$\omega_{\mathrm{b}}$ is the amplitude of longitudinal spin driving [see Eq.
(\ref{omega_0})], the dimensionless coupling coefficient $\beta_{\Delta}$ is
given by $\beta_{\Delta}=\omega_{\Delta}/\sqrt{\gamma_{1}\gamma_{2}}$, the
dimensionless detuning coefficients $\beta_{\mathrm{c}l}$ and $\beta
_{\mathrm{a}l}$ are given by $\beta_{\mathrm{c}l}=\left(  \omega_{\mathrm{c}%
}-l\omega_{\mathrm{LA}}\right)  /\gamma_{\mathrm{c}}$ and $\beta_{\mathrm{a}%
l}=\left(  \omega_{\mathrm{a}}-l\omega_{\mathrm{LA}}\right)  /\gamma_{2}$,
respectively, $\omega_{\mathrm{c}}$ is the cavity mode angular frequency,
$\gamma_{\mathrm{c}}$ is the cavity mode damping rate, $\gamma_{1}$ and
$\gamma_{2}$ are the longitudinal and transverse spin damping rates,
respectively and $\kappa=g^{2}/\gamma_{2}\gamma_{\mathrm{c}}$ is the
cooperativity parameter. A plot of the normalized steady state polarization
$P_{z}/P_{z,\mathrm{s}}$ given by Eq. (\ref{P_z SS=}) is shown in Fig.
\ref{Fig_LowL}(b). The comparison between data and theory yields a qualitative agreement. 

\section{P1}

\begin{figure}
[ptb]
\begin{center}
\includegraphics[
height=2.5949in,
width=3.4537in
]%
{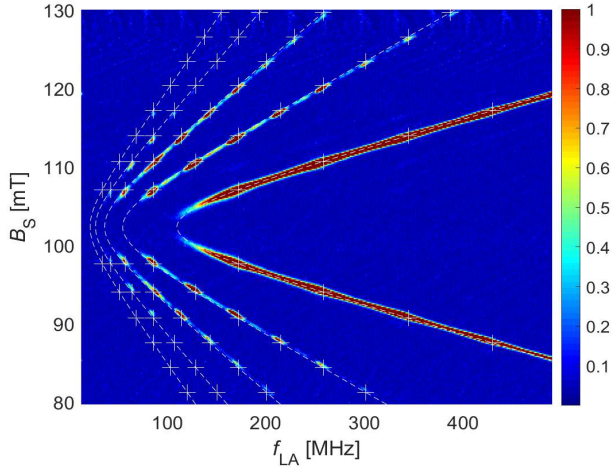}%
%ParamSet 102 (data 20180108-1801)
\caption{ODMR with high laser power. The overlaid hyperbolas are calculated
according to Eq. (\ref{hyperbolas}) and the locations of the cross symbols
according to Eq. (\ref{f_k,l}).}%
\label{Fig_HighL}%
\end{center}
\end{figure}

ODMR data near the LAC with relatively high laser power is shown in
Fig. \ref{Fig_HighL}. The increase in laser power gives rise to excessive heating, and consequently the superconducting resonator mode becomes undetectable (in a microwave reflectivity measurement) due to a super to normal conduction phase transition of the spiral. The plot contains a variety of
peaks all occurring along the above discussed hyperbolas [see Eq. (\ref{hyperbolas})], suggesting that some multiphoton processes continue to exist regardless of the spiral resonator state. Locations of all data peaks are determined by a single frequency denoted by $f_{\mathrm{m}}$.
This can be seen from the cross symbols added to Fig. \ref{Fig_HighL}. The
frequency $f_{k,l}$ of the $k$'th cross symbol overlaid on the $l$'th
hyperbola in Fig. \ref{Fig_HighL} is given by%
\begin{equation}
f_{k,l}=\frac{k}{l}f_{\mathrm{m}}\;,\label{f_k,l}%
\end{equation}
where the frequency $f_{\mathrm{m}}$ takes the value $f_{\mathrm{m}}=86%
%TCIMACRO{\unit{MHz}}%
%BeginExpansion
\operatorname{MHz}%
%EndExpansion
$. 
This pattern of peaks remains visible with the same value of $f_{\mathrm{m}}$  over a wide range of input microwave power (between $10$ and $25\operatorname{dBm}$), tenfold laser power attenuation, a few degrees magnetic field misalignments and temperature change. With temperature rising to $30\operatorname{K}$, the signal from the higher order hyperbolas disappears, but the $f_\mathrm{m}$ beating remains on the main hyperbola. The fact that some of the peaks do not appear at the same frequency for different magnetic fields validates that the pattern is not a measurement artifact of spurious resonances. In addition, the synthesizer signal harmonics were carefully examined with a spectrum analyzer to verify they are all well below the ODMR sensitivity threshold.
The measured value of $f_{\mathrm{m}}$ suggests a connection between MPR in the NV$^{-}$ defects and P1 defect \cite{Smith_1546,Cook_99,Loubser_1201,Barklie_3621}, as is discussed below.%

%TCIMACRO{\FRAME{ftbpFU}{3.4537in}{2.5949in}{0pt}{\Qcb{ODMR with high laser
%power. The overlaid hyperbolas are calculated according to Eq.
%(\ref{hyperbolas}) and the locations of the cross symbols according to Eq.
%(\ref{f_k,l}).}}{\Qlb{Fig_HighL}}{fig_highl.eps}%
%maintain-aspect-ratio TRUE;  display "ICON";  valid_file "F";
%width 3.4537in;  height 2.5949in;  depth 0pt;  original-width 8.5365in;
%original-height 6.3945in;  cropleft "0";  croptop "1";  cropright "1";
%cropbottom "0";  filename 'Fig_HighL.eps';file-properties "XNPEU";}} }%
%BeginExpansion
%EndExpansion

P1 defect has four locally stable
configurations. In each configuration a static Jahn-Teller distortion occurs,
and an unpaired electron is shared by the nitrogen atom and by one of the four
neighboring carbon atoms, which are positioned along one of the lattice
directions $\left[  111\right]  $, $\left[  \bar{1}11\right]  $, $\left[
1\bar{1}1\right]  $ or $\left[  11\bar{1}\right]  $
\cite{Takahashi_047601,Hanson_352,Wang_4135,Broadway_1607_04006,Shim_1307_0257,Smeltzer_025021,Shin_205202,Clevenson_021401,Schuster_140501}%
.

When both nuclear Zeeman shift and nuclear quadrupole coupling are
disregarded, the spin Hamiltonian of a P1 defect is given by
\cite{Loubser_1201,Schuster_140501,Wood_155402} $\mathcal{H}=\gamma
_{\mathrm{e}}\mathbf{B}\cdot\mathbf{S}+\hbar^{-1}A_{\perp}\left(  S_{x}%
I_{x}+S_{y}I_{y}\right)  +\hbar^{-1}A_{\parallel}S_{z}I_{z}$, where
$\mathbf{S}=\left(  S_{x},S_{y},S_{z}\right)  $ is an electronic spin 1/2
vector operator, $\mathbf{I}=\left(  I_{x},I_{y},I_{z}\right)  $ is a nuclear
spin 1 vector operator, $A_{\parallel}=2\pi\times114.03%
%TCIMACRO{\unit{MHz}}%
%BeginExpansion
\operatorname{MHz}%
%EndExpansion
$ and $A_{\perp}=2\pi\times81.33%
%TCIMACRO{\unit{MHz}}%
%BeginExpansion
\operatorname{MHz}%
%EndExpansion
$ are respectively the longitudinal and transverse hyperfine parameters, and
the $z$ direction corresponds to the diamond $\left\langle 111\right\rangle $
axis. The electron spin resonance at angular frequency $\gamma_{\mathrm{e}}B$
is split due to the interaction with the nuclear spin into three resonances,
corresponding to three transitions, in which the nuclear spin magnetic quantum
number is conserved
\cite{Slichter_Principles,Simanovskaia_224106,Belthangady_157601,vanOort_8605,Hanson_352,Hall_10211}%
. For a magnetic field larger than a few $%
%TCIMACRO{\unit{mT}}%
%BeginExpansion
\operatorname{mT}%
%EndExpansion
$ the angular resonance frequencies are approximately given by $\gamma
_{\mathrm{e}}B$ and $\gamma_{\mathrm{e}}B\pm\omega_{\mathrm{en}}$, where
$\omega_{\mathrm{en}}^{2}=A_{\parallel}^{2}\cos^{2}\theta_{\mathrm{B}%
}+A_{\perp}^{2}\sin^{2}\theta_{\mathrm{B}}$ and where $\theta_{\mathrm{B}}$ is
the angle between the magnetic field $\mathbf{B}$ and the P1 axis
\cite{Smith_1546}.

Consider the case where $\mathbf{B}$ is in the lattice direction $\left[
111\right]  $. For this case, for 1/4 of the P1 defects $\omega_{\mathrm{en}%
}=2\pi\times114%
%TCIMACRO{\unit{MHz}}%
%BeginExpansion
\operatorname{MHz}%
%EndExpansion
$, whereas for the other 3/4 of the P1 defects [unparallel to $\mathbf{B}$
having axis in one of the lattice directions $\left[  \bar{1}11\right]  $,
$\left[  1\bar{1}1\right]  $ or $\left[  11\bar{1}\right]  $] $\omega
_{\mathrm{en}}=2\pi\times85.6\operatorname{MHz}$, close to the observed value
of the frequency $f_{\mathrm{m}}=86\operatorname{MHz}$. The fact that the parallel
P1 defects do not have a significant effect on the ODMR data can be attributed
to the fact that these defects generate only transverse driving for the
NV$^{-}$ defects having axis parallel to the crystal direction $\left[
111\right]  $, whereas the unparallel P1 defects generate both transverse and
longitudinal driving, which in turn allows nonlinear processes of frequency mixing \cite{Cohen_atom}.

%The angular frequency of nuclear-like transitions of unparallel P1 defects is
%$\left(  \omega_{\mathrm{en}}/2\right)  /2\pi=42.8%
%%TCIMACRO{\unit{MHz}}%
%%BeginExpansion
%\operatorname{MHz}%
%%EndExpansion
%$. Such an halved spacing, corresponding to nuclear-like transitions with
%$\Delta m_{\mathrm{n}}=\pm1$, is experimentally observed as weak peaks on the
%second (i.e. $l=2$) hyperbola in Fig. \ref{Fig_HighL}. However, the strong
%peaks on this $l=2$\ hyperbola as well as all other peaks on the other
%hyperbolas are well described by Eq. (\ref{f_k,l}) with the frequency
%$f_{\mathrm{m}}=86\operatorname{MHz}$.

The effect of dipolar interactions on the measured ODMR signal can be estimated using perturbation theory. To first order the above-discussed hyperfine splitting has no effect. However, as is argued below, a non-vanishing effect is obtained from the second order. Consider a pair of P1 defects with a dipolar coupling to a single NV$^{-}$ defect \cite{Hanson_087601,Bermudez_150503,Kohl_319,Cheng_1359,Daycock_998,Alekseev_1272,Varada_6721}. Both P1 defects
are assumed to be unparallel to $\mathbf{B}$, i.e. the frequencies of their
electronic-like transitions are approximately given by $\gamma_{\mathrm{e}}B/2\pi$ and $\gamma_{\mathrm{e}}B/2\pi\pm85.6\operatorname{MHz}$. The NV$^{-}$ defect, on the other hand, is assumed to be nearly parallel to
$\mathbf{B}$, thus having an energy separation of $2\hbar\gamma_{\mathrm{e}}B$ between the spin states with magnetic number $\pm1$. %Forbidden at a perfect alignment of the NV defect along $\mathbf{B}$, transition between these states is allowed by the tripolar coupling.

OISP polarizes the NV$^-$ to the $m_\mathrm{s}=0$ state. The required condition for ODMR signal along the $l$-th hyperbola is achieved by excitation at  $\omega_\mathrm{LA}=\omega_\mathrm{a}/l$, populating the $m_\mathrm{s}=-1$ state, which has lower photoluminescence.
% with $i\in\{+1,0,1\}$ and . 
Let $(m_\mathrm{S}^\mathrm{NV},m_\mathrm{S}^\mathrm{P1a}+m_\mathrm{S}^\mathrm{P1b},m_\mathrm{I}^\mathrm{P1a}+m_\mathrm{I}^\mathrm{P1b})$ designate a subspace, where $m_\mathrm{S}^\mathrm{NV}$ is the NV$^-$ electronic spin magnetic number, $m_\mathrm{S}^\mathrm{P1a}$ ($m_\mathrm{S}^\mathrm{P1b}$) is the first (second) P1 electronic spin magnetic number and $m_\mathrm{I}^\mathrm{P1a}$ ($m_\mathrm{I}^\mathrm{P1b}$) is the first (second) P1 nuclear spin magnetic number.
Note that subspaces $(-1,+1,j)$ and $(+1,-1,j)$ for $j\in \{-2,-1,0,1,2\}$, are energetically separated by $\hbar|j|\omega_\mathrm{en}$. When $\omega_\mathrm{a}=k\omega_\mathrm{en}$ for integer $k$, transitions are stimulated between $(-1,+1,j)$ and $(+1,-1,j)$, further reducing the population of $(0,+1,j)$ and consequently enhancing the ODMR signal. 

By employing perturbation theory
\cite{Schrieffer_491} we find that the effective Rabi rate for these tripolar transitions
is roughly given by $\omega_{\mathrm{P1P1NV}}\simeq\left(  n_{\mathrm{S}%
	,\mathrm{P1}}/n_{\mathrm{D}}\right)  ^{2}D$, where $n_{\mathrm{S},\mathrm{P1}%
}$ is the density of P1 defects (which is assumed to be about $100$ times larger than the
density of NV$^{-}$ defects, and which can be expressed in terms of the
relative concentration of nitrogen atoms $p_{\mathrm{N}}$ as $n_{\mathrm{S}%
	,\mathrm{P1}}=1.8\times10^{23}%
%TCIMACRO{\unit{cm}}%
%BeginExpansion
\operatorname{cm}%
%EndExpansion
^{-3}p_{\mathrm{N}}$) and where $n_{\mathrm{D}}=4\pi D/\mu_{0}\gamma
_{\mathrm{e}}^{2}\hbar=5.5\times10^{22}%
%TCIMACRO{\unit{cm}}%
%BeginExpansion
\operatorname{cm}%
%EndExpansion
^{-3}$. The roughly estimated value of $p_{\mathrm{N}}=10^{-4}$ yields the
rate $\omega_{\mathrm{P1P1NV}}/2\pi\simeq300%
%TCIMACRO{\unit{kHz}}%
%BeginExpansion
\operatorname{Hz}%
%EndExpansion
$. 
%The effective rate is further enhanced by the large number of possible NV-P1-P1 triples, i.e no transitions are possible for $(\pm1,0,j)$, but for a different choice of P1 the relevant subspace will become $(\pm1,\mp1,j)$, allowing the NV$^-$ $m_\mathrm{s}=+1\leftrightarrow-1$ transition. 
In a similar setup \cite{Alfasi_2018}, at $T=3.5\operatorname{K}$ the maximal OISP rate was found to be $T_\mathrm{1O}^{-1}\approx 200\operatorname{Hz}$ and $\gamma_1\approx 25\operatorname{Hz}$, hence this mechanism is expected to be of significance to a low temperature ODMR measurement. 

Note that the transition between $(-1,+1,0)$ and $(+1,-1,0)$ does not require additional energy. 
Stimulated nuclear spin rotation with $\omega_\mathrm{a}=k\omega_\mathrm{en}/2$ for integer $k$ allows population of $(+1,-1,j_1)$ for $j_1\in\{-2,-1,1,2\}$ via processes of sequential photons absorption. This effect gives rise to the weak peaks on the
second ($l=2$) hyperbola in Fig. \ref{Fig_HighL}.  %Another mechanism that allows direct ODMR measurement of the P1 electronic spin state is the inverse to hyperpolarization \cite{Alfasi_2018}, namely : P1  are first optically hyperoplarized due to a dipolar coupling to NV, but when their polarization is reduced by microwave photons absorption, the polarization of NV is reduced as well due to the same dipolar coupling.
%account for the
%enhancement of multi-photon resonances at frequencies $f_{k,l}=\left(
%k/l\right)  f_{\mathrm{m}}$, where $f_{\mathrm{m}}=86%
%%TCIMACRO{\unit{MHz}}%
%%BeginExpansion
%\operatorname{MHz}%
%%EndExpansion
%$, which is experimentally observed near the avoided crossing point (see Fig.
%\ref{Fig_HighL}). Note that, in addition, the induced transitions between the
%states $\left\vert -1/2,-1/2,1\right\rangle $ and $\left\vert
%1/2,1/2,-1\right\rangle $ are expected to give rise to depolarization of both
%P1 and NV$^{-}$ defects.

\begin{figure}
%	\begin{center}
		\includegraphics[width=\columnwidth]{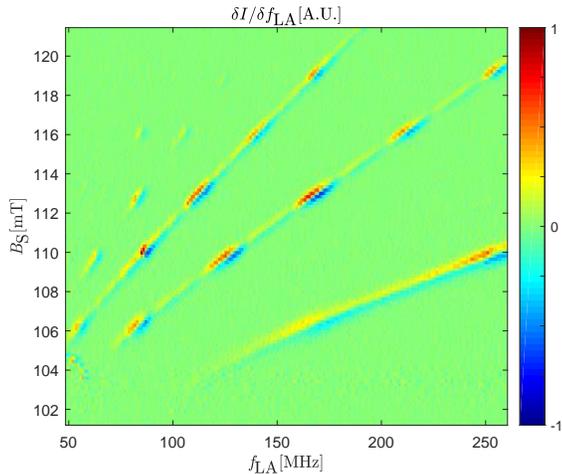}
		\caption{Derivative of the ODMR photoluminescence signal $I$ with respect to the driving signal frequency $f_\textmd{LA}$ at high laser power. The absolute maximal values are achieved not along the single photon curve (bottom right diagonal), but rather at the spots attributed to P1 hyperfine processes on the NV MPR curves (top left diagonals).}
		\label{Fig_dIdf}
%	\end{center}
\end{figure}

The MPR can be employed for enhancing the responsivity of diamond based magnetometry. Consider a setup with small frequency modulation about a central frequency $f_\textmd{LA}$ and photoluminescence signal $I$ demodulation readout. To maximize the responsivity, the bias magnetic field $B_\textmd{S}$ and $f_\textmd{LA}$ should be set to maximize the derivative $|\mathrm{d}I / \mathrm{d}f_\textmd{LA}|$. As can be seen in Fig.~\ref{Fig_dIdf}, $|\mathrm{d}I / \mathrm{d}f_\textmd{LA}|$ is maximal near the spots associated with P1 hyperfine transitions at the MPR of NV. This enhancement is attributed to the relatively narrow resonance of the P1 process as compared to the NV MPR.
%Additional applications for the above discussed mechanisms of electronic spin exchange between a well oriented NV$^-$ defect to unperturbed electronic spins in the environment are spins detection and hyperpolarization. 

\section{Summary}
Multiphoton processes are surprisingly well measurable in Ib diamonds, making this mode of operation preferable for enhanced sensitivity in multiple applications. Of particular interest is the interaction of the optically measurable and polarizable NV ensemble with the naturally occurring P1 ensemble. The unexpected strength of coupling to the hyperfine transitions of the P1 requires further investigation to determine the nature of the interaction. The NV defects can potentially provide an optical access to a much denser and coherent (nuclear) ensemble of the P1.

\section{Acknowledgements}

We greatly appreciate fruitful discussions with Paz London, Aharon Blank, Efrat Lifshitz, Vladimir Dyakonov, Sergey Tarasenko, Victor Soltamov, Nadav Katz, Michael Stern and Nir Bar-Gil.

\appendix

\section{Driven Spins Coupled to a Resonator}
\label{appA}
Consider a cavity mode coupled to a spin ensemble. The Hamiltonian of the closed system is taken to be given by%
\begin{equation}
\hbar ^{-1}\mathcal{H}_{0}=\omega _{\mathrm{c}}A^{\dag }A+\frac{\mathbf{%
		\Omega \cdot \Sigma }}{2}+g\left( A\Sigma _{+}+A^{\dag }\Sigma _{-}\right) \
,  \label{H_0 Sigma}
\end{equation}%
where $\omega _{\mathrm{c}}$ is the cavity mode angular frequency, $A^{\dag }A$ is a cavity mode number operator, $\mathbf{\Sigma }=\left( \Sigma _{x},\Sigma _{y},\Sigma
_{z}\right) $, the spin operators $\Sigma _{z}$, $\Sigma _{+}=\left( \Sigma
_{x}+i\Sigma _{y}\right) /2$ and $\Sigma _{-}=\left( \Sigma _{x}-i\Sigma
_{y}\right) /2$ are related to the eigenvectors $\left\vert \pm
\right\rangle $ of the operator $\Sigma_{z} $ by%
\begin{align}
\Sigma _{z}& =\left\vert +\right\rangle \left\langle +\right\vert
-\left\vert -\right\rangle \left\langle -\right\vert \;, \\
\Sigma _{+}& =\left\vert +\right\rangle \left\langle -\right\vert \;, \\
\Sigma _{-}& =\left\vert -\right\rangle \left\langle +\right\vert \;.
\end{align}%
The effective magnetic field $\mathbf{\Omega }\left(
t\right) $ is expressed in terms of the angular frequency $\omega _{\mathrm{T%
}}$ and amplitude $\omega _{1}$ of transverse driving, the longitudinal
magnetic field component $\omega _{0}\left( t\right) $ and transverse one $%
\omega _{\Delta }$%
\begin{equation}
\mathbf{\Omega }\left( t\right) =\omega _{1}\left( \cos \left( \omega _{%
	\mathrm{T}}t\right) \mathbf{\hat{x}}+\sin \left( \omega _{\mathrm{T}%
}t\right) \mathbf{\hat{y}}\right) +\omega _{0}\mathbf{\hat{z}}+\omega
_{\Delta }\mathbf{\hat{x}}\ ,
\end{equation}%
or%
\begin{align}
\mathbf{\Omega }\left( t\right) & =\omega _{1}\left( e^{-i\omega _{\mathrm{T}%
	}t}\mathbf{\hat{u}}_{+}+e^{i\omega _{\mathrm{T}}t}\mathbf{\hat{u}}%
_{-}\right) +\omega _{0}\left( t\right) \mathbf{\hat{z}}  \notag \\
& +\omega _{\Delta }\left( \mathbf{\hat{u}}_{+}+\mathbf{\hat{u}}_{-}\right)
\ ,  \notag \\
&  \label{Omega(t)}
\end{align}%
where $\omega _{\Delta }$ is a real constant and%
\begin{equation}
\mathbf{\hat{u}}_{\pm }=\left( 1/2\right) \left( \mathbf{\hat{x}}\pm i%
\mathbf{\hat{y}}\right) \ .
\end{equation}%
While $\omega _{1}$ and $\omega _{\mathrm{T}}$ are both assumed to be real
constants, $\omega _{0}$ is allowed to vary in time according to%
\begin{equation}
\omega _{0}=\omega _{\mathrm{a}}-\omega _{\mathrm{b}}\sin \left( \omega _{%
	\mathrm{L}}t\right) \;,  \label{omega_0}
\end{equation}%
where $\omega _{\mathrm{a}}$, $\omega _{\mathrm{b}}$ and $\omega _{\mathrm{L}%
}$ are all real constants.

The following Bose%
\begin{equation}
\left[ A,A^{\dagger}\right] =1\;,
\end{equation}
and spin%
\begin{align}
\left[ \Sigma_{z},\Sigma_{+}\right] & =2\Sigma_{+}\;,  \label{CR z+} \\
\left[ \Sigma_{z},\Sigma_{-}\right] & =-2\Sigma_{-}\;,  \label{CR z-} \\
\left[ \Sigma_{+},\Sigma_{-}\right] & =\Sigma_{z}\;,  \label{CR +-}
\end{align}
commutation relations are assumed to hold. The Heisenberg equations of
motion are generated according to 
\begin{equation}
\frac{\mathrm{d}O}{\mathrm{d}t}=-i\left[ O,\hbar^{-1}\mathcal{H}_{0}\right]
\;,  \label{Heisenberg}
\end{equation}
where $O$ is an operator, hence%
\begin{equation}
\frac{\mathrm{d}A}{\mathrm{d}t}=-i\omega_{\mathrm{c}}A-ig\Sigma_{-}\;,
\end{equation}%
\begin{equation}
\frac{\mathrm{d}\Sigma_{z}}{\mathrm{d}t}=W_{1}\Sigma_{+}+W_{1}^{\dag}%
\Sigma_{-}\;,
\end{equation}
and%
\begin{equation}
\frac{\mathrm{d}\Sigma_{+}}{\mathrm{d}t}=i\omega_{0}\Sigma_{+}-\frac {%
	W_{1}^{\dag}}{2}\Sigma_{z}\;,
\end{equation}
where%
\begin{equation}
W_{1}=-i\left( \omega_{1}e^{-i\omega_{\mathrm{T}}t}+\omega_{\Delta
}+2gA\right) \;.
\end{equation}

Averaging%
\begin{align}
\left\langle A\right\rangle & =\alpha \;, \\
\left\langle \Sigma _{z}\right\rangle & =P_{z}\;, \\
\left\langle \Sigma _{\pm }\right\rangle & =P_{\pm }\;,
\end{align}%
and introducing damping leads to%
\begin{equation}
\frac{\mathrm{d}\alpha }{\mathrm{d}t}=-\left( i\omega _{\mathrm{c}}+\gamma _{%
	\mathrm{c}}\right) \alpha -igP_{-}\;,  \label{d alpha/dt V1}
\end{equation}%
\begin{equation}
\frac{\mathrm{d}P_{z}}{\mathrm{d}t}=\Omega _{1}P_{+}+\Omega _{1}^{\ast
}P_{-}-\gamma _{1}\left( P_{z}-P_{z,\mathrm{s}}\right) \;,
\label{dP_z/dt V1}
\end{equation}%
and%
\begin{equation}
\frac{\mathrm{d}P_{+}}{\mathrm{d}t}=i\omega _{0}P_{+}-\frac{\Omega
	_{1}^{\ast }}{2}P_{z}-\gamma _{2}P_{+}\;,  \label{dP_+/dt V1}
\end{equation}%
where $\gamma _{\mathrm{c}}$ is the cavity mode damping rate, $\gamma _{1}$
and $\gamma _{2}$ are the longitudinal and transverse spin damping rates,
respectively, and where%
\begin{equation}
\Omega _{1}=-i\left( \omega _{1}e^{-i\omega _{\mathrm{T}}t}+\omega _{\Delta
}+2g\alpha \right) \;.  \label{Omega_1}
\end{equation}%
For our experimental conditions the term proportional to $\omega _{1}$ in
Eq. (\ref{Omega_1}) can be disregarded.

The effect of OISP can be accounted for by adjusting the values of the
longitudinal damping rate $\gamma _{1}$ and steady state polarization $P_{z,%
	\mathrm{s}}$ and make them both dependent on laser intensity \cite%
{Shin_124519,Alfasi_063808}. In this approach $\gamma _{1}$ is given by $%
\gamma _{1}=\gamma _{1\mathrm{T}}+\gamma _{1\mathrm{O}}$, where $\gamma _{1%
	\mathrm{T}}$ is the rate of thermal relaxation and $\gamma _{1\mathrm{O}}$
is the rate of OISP (proportional to laser intensity), and the averaged
value of steady state polarization $P_{z,\mathrm{s}}$ is given by%
\begin{equation}
P_{z,\mathrm{s}}=\frac{\gamma _{1\mathrm{T}}P_{z,\mathrm{ST}}+\gamma _{1%
		\mathrm{O}}P_{z,\mathrm{SO}}}{\gamma _{1}}\;.
\end{equation}%
While $P_{z,\mathrm{ST}}$ represents the steady state polarization in the
limit $\gamma _{1\mathrm{T}}\gg \gamma _{1\mathrm{O}}$ (i.e. when OISP is
negligibly small), the value is $P_{z,\mathrm{SO}}$ for the other extreme
case of $\gamma _{1\mathrm{O}}\gg \gamma _{1\mathrm{T}}$ (i.e. when thermal
relaxation is negligibly small).

By employing the transformation%
\begin{equation}
P_{+}=e^{i\theta _{\mathrm{d}}}P_{\mathrm{d}+}\ ,  \label{P_+}
\end{equation}%
where%
\begin{equation}
\theta _{\mathrm{d}}=\int\nolimits^{t}\mathrm{d}t^{\prime }\;\left[ \omega
_{0}\left( t^{\prime }\right) +\Delta \right] \;,  \label{theta_d}
\end{equation}%
and where $\Delta $ is a real constant (to be determined later), Eqs. (\ref%
{d alpha/dt V1}), (\ref{dP_z/dt V1}) and (\ref{dP_+/dt V1}) become%
\begin{equation}
\frac{\mathrm{d}\alpha }{\mathrm{d}t}=-\left( i\omega _{\mathrm{c}}+\gamma _{%
	\mathrm{c}}\right) \alpha -ig\left( \frac{\omega _{\Delta }\zeta }{\Omega
	_{1}}\right) ^{\ast }P_{\mathrm{d}+}^{\ast }\;,  \label{d alpha/dt V2}
\end{equation}%
\begin{equation}
\frac{\mathrm{d}P_{z}}{\mathrm{d}t}=\omega _{\Delta }\left( \zeta P_{\mathrm{%
		d}+}+\zeta ^{\ast }P_{\mathrm{d}+}^{\ast }\right) -\gamma _{1}\left(
P_{z}-P_{z,\mathrm{s}}\right) \;,  \label{dP_z/dt V2}
\end{equation}%
and%
\begin{equation}
\frac{\mathrm{d}P_{\mathrm{d}+}}{\mathrm{d}t}=-i\Delta P_{\mathrm{d}+}-\frac{%
	\omega _{\Delta }\zeta ^{\ast }}{2}P_{z}-\gamma _{2}P_{\mathrm{d}+}\;,
\label{dP_+/dt V2}
\end{equation}%
where%
\begin{equation}
\zeta =\frac{\Omega _{1}}{\omega _{\Delta }}e^{i\theta _{\mathrm{d}}}\;.
\label{zeta}
\end{equation}%
When $\zeta $ is treated as a constant the steady state solution of Eqs. (%
\ref{dP_z/dt V2}) and (\ref{dP_+/dt V2}) reads%
\begin{equation}
P_{\mathrm{d}+}=\frac{\omega _{\Delta }\zeta ^{\ast }P_{z}}{2\left( -i\Delta
	-\gamma _{2}\right) }\;,  \label{P_d+ SS}
\end{equation}%
and%
\begin{equation}
\frac{P_{z}}{P_{z,\mathrm{s}}}=1-\frac{\frac{\left\vert \omega _{\Delta
		}\zeta \right\vert ^{2}}{\gamma _{1}\gamma _{2}}}{1+\frac{\left\vert \omega
		_{\Delta }\zeta \right\vert ^{2}}{\gamma _{1}\gamma _{2}}+\frac{\Delta ^{2}}{%
		\gamma _{2}^{2}}}\;.  \label{P_z SS}
\end{equation}

With the help of the Jacobi-Anger expansion, which is given by%
\begin{equation}
\exp\left( iz\cos\theta\right)
=\sum\limits_{n=-\infty}^{\infty}i^{n}J_{n}\left( z\right) e^{in\theta}\;,
\label{Jacobi-Anger}
\end{equation}
one obtains [see Eqs. (\ref{omega_0}) and (\ref{zeta})]%
\begin{equation}
\zeta=\frac{\Omega_{1}}{\omega_{\Delta}}e^{-\frac{i\omega_{\mathrm{b}}}{%
		\omega_{\mathrm{L}}}}\sum\limits_{l^{\prime}=-\infty}^{\infty}i^{l^{%
		\prime}}J_{l^{\prime}}\left( \frac{\omega_{\mathrm{b}}}{\omega_{\mathrm{L}}}%
\right) e^{i\left( \omega_{\mathrm{a}}+\Delta+l^{\prime}\omega_{\mathrm{L}%
	}\right) t}\;.  \label{zeta V1}
\end{equation}
Consider the case where $\omega_{\mathrm{a}}\simeq l\omega_{\mathrm{L}}$,
where $l$ is an integer. For this case the detuning $\Delta$ is chosen to be
given by $\Delta=l\omega_{\mathrm{L}}-\omega_{\mathrm{a}}$, and consequently 
$\zeta$ becomes%
\begin{equation}
\zeta=\frac{\Omega_{1}}{\omega_{\Delta}}e^{-\frac{i\omega_{\mathrm{b}}}{%
		\omega_{\mathrm{L}}}}\sum\limits_{l^{\prime}=-\infty}^{\infty}i^{l^{%
		\prime}}J_{l^{\prime}}\left( \frac{\omega_{\mathrm{b}}}{\omega_{\mathrm{L}}}%
\right) e^{i\left( l+l^{\prime}\right) \omega_{\mathrm{L}}t}\;.
\label{zeta V2}
\end{equation}

The driving term of Eq. (\ref{d alpha/dt V2}) $-ig\left(
\omega_{\Delta}\zeta/\Omega_{1}\right) ^{\ast}P_{\mathrm{d}+}^{\ast}$ is
approximated by keeping only the term $l^{\prime}=0$ in Eq. (\ref{zeta V2}).
When $P_{\mathrm{d}+}^{\ast}$ is treated as a constant Eq. (\ref{d alpha/dt
	V2}) yields a steady state solution given by $\alpha=\alpha_{0}e^{-il\omega
	_{\mathrm{L}}t}$, where%
\begin{equation}
\alpha_{0}=-\frac{ige^{\frac{i\omega_{\mathrm{b}}}{\omega_{\mathrm{L}}}%
	}J_{0}\left( \frac{\omega_{\mathrm{b}}}{\omega_{\mathrm{L}}}\right) P_{%
		\mathrm{d}+}^{\ast}}{\gamma_{\mathrm{c}}\left( 1+i\beta_{\mathrm{c}l}\right) 
}\;,  \label{alpha_0}
\end{equation}
and where%
\begin{equation}
\beta_{\mathrm{c}l}=\frac{\omega_{\mathrm{c}}-l\omega_{\mathrm{L}}}{\gamma_{%
		\mathrm{c}}}\;.  \label{beta_cl}
\end{equation}
To lowest non vanishing order in the coupling $g$ the coefficient $P_{%
	\mathrm{d}+}^{\ast}$ in Eq. (\ref{alpha_0}) is evaluated using Eq. (\ref%
{P_d+ SS}) by keeping only the term $l^{\prime}=-l$ in Eq. (\ref{zeta V2})
and keeping only the term $-i\omega_{\Delta}$ in Eq. (\ref{Omega_1})%
\begin{equation}
P_{\mathrm{d}+}^{\ast}=\frac{i^{1-l}e^{-\frac{i\omega_{\mathrm{b}}}{\omega_{%
				\mathrm{L}}}}\omega_{\Delta}J_{-l}\left( \frac{\omega_{\mathrm{b}}}{\omega_{%
			\mathrm{L}}}\right) P_{z,\mathrm{s}}}{2\gamma_{2}\left( 1+i\beta_{\mathrm{a}%
		l}\right) }\;,
\end{equation}
where%
\begin{equation}
\beta_{\mathrm{a}l}=\frac{\omega_{\mathrm{a}}-l\omega_{\mathrm{L}}}{\gamma
	_{2}}\;,  \label{beta_al}
\end{equation}
and thus [see Eq. (\ref{alpha_0})]%
\begin{equation}
\alpha_{0}=-\frac{i^{2-l}g\omega_{\Delta}}{\gamma_{\mathrm{c}}\gamma_{2}}%
\frac{J_{0}\left( \frac{\omega_{\mathrm{b}}}{\omega_{\mathrm{L}}}\right)
	J_{-l}\left( \frac{\omega_{\mathrm{b}}}{\omega_{\mathrm{L}}}\right) P_{z,%
		\mathrm{s}}}{2\left( 1+i\beta_{\mathrm{c}l}\right) \left( 1+i\beta_{\mathrm{a%
		}l}\right) }\;.  \label{alpha_0 V1}
\end{equation}

It is assumed that the dominant contribution of $\zeta $ to the equation of
motion (\ref{dP_z/dt V2}) and (\ref{dP_+/dt V2}) comes from a term, which is
labelled as $\zeta _{\mathrm{a}}$, which is given by [see Eqs. (\ref{Omega_1}%
) and (\ref{zeta V2})]%
\begin{equation}
ie^{\frac{i\omega _{\mathrm{b}}}{\omega _{\mathrm{L}}}}\zeta _{\mathrm{a}%
}=i^{-l}J_{-l}\left( \frac{\omega _{\mathrm{b}}}{\omega _{\mathrm{L}}}%
\right) +\frac{2g\alpha _{0}}{\omega _{\Delta }}J_{0}\left( \frac{\omega _{%
		\mathrm{b}}}{\omega _{\mathrm{L}}}\right) \;.
\end{equation}%
With the help of Eq. (\ref{alpha_0 V1}) this becomes%
\begin{equation}
i^{1+l}e^{\frac{i\omega _{\mathrm{b}}}{\omega _{\mathrm{L}}}}\zeta _{\mathrm{%
		a}}=J_{-l}\left( \frac{\omega _{\mathrm{b}}}{\omega _{\mathrm{L}}}\right)
\left( 1+\frac{\kappa J_{0}^{2}\left( \frac{\omega _{\mathrm{b}}}{\omega _{%
			\mathrm{L}}}\right) P_{z,\mathrm{s}}}{\left( 1+i\beta _{\mathrm{c}l}\right)
	\left( 1+i\beta _{\mathrm{a}l}\right) }\right) \;,  \label{zeta_a}
\end{equation}%
where the cooperativity parameter $\kappa $ is given by%
\begin{equation}
\kappa =\frac{g^{2}}{\gamma _{2}\gamma _{\mathrm{c}}}\;.
\end{equation}%
The above results (\ref{P_z SS}) and (\ref{zeta_a}) lead to Eq. (\ref{P_z
	SS=}) in main text for the steady state polarization.

\newpage
%Just because of unusual number of tables stacked at end
%\bibliographystyle{ieee}
%\bibliography{acompat,Eyal_Bib}

\bibliography{Eyal_Bib}%Produces the bibliography via BibTeX.

\end{document}